\documentclass{article}
\usepackage[a4paper, width=170mm, top=25mm, bottom=25mm, bindingoffset=6mm]{geometry} 
\usepackage{authblk} 

\usepackage{graphicx} 
\usepackage{subcaption} 
\usepackage{float} 
\usepackage{comment}

\usepackage{amsmath} 
\usepackage{enumitem} 
\usepackage{amssymb} 

\usepackage[backend=biber,style=numeric, sorting=none]{biblatex} 
\usepackage[colorlinks=true,linkcolor=blue,citecolor=blue,urlcolor=blue]{hyperref}

\usepackage{xcolor}

\title{Ideological polarization in static networks: A multidimensional approach for opinion alignment}

\author[a,b]{Favio Di Ciocco}
\author[c,d]{Hugo P\'erez-Mart\'inez}
\author[c,d]{Jes\'us G\'omez-Garde\~nes}
\author[e,c]{David Soriano-Pa\~nos}
\author[a,b]{Pablo Balenzuela}

\affil[a]{Universidad de Buenos Aires, Facultad de Ciencias Exactas y Naturales, Departamento de F\'isica. Buenos Aires, Argentina}
\affil[b]{CONICET - Universidad de Buenos Aires, Instituto de F\'isica Interdisciplinaria y Aplicada (INFINA), Ciudad Universitaria, 1428. Buenos Aires, Argentina}
\affil[c]{GOTHAM lab, Institute for Biocomputation and Physics of Complex Systems (BIFI), University of Zaragoza, 50018, Zaragoza, Spain}
\affil[d]{Department of Condensed Matter Physics, University of Zaragoza, 50009, Zaragoza, Spain}
\affil[e]{Departament d'Enginyeria Inform{\`a}tica i Matem{\`a}tiques,Universitat Rovira i Virgili, 43007 Tarragona, Spain.}

\date{\today}

\begin{document}

\maketitle

\begin{abstract}

Polarization—characterized by the formation of sharply divided groups holding opposing and often extreme views—has become an increasingly prominent feature of contemporary societies. While numerous studies have examined this phenomenon through the lens of single-issue dynamics, such as public opinion on abortion or immigration, this approach overlooks a critical aspect of real-world discourse: political and social attitudes rarely develop in isolation. Instead, many of these issues are deeply interconnected, shaped by overarching ideological frameworks that guide individuals’ interpretations and position-taking across multiple topics. These frameworks give rise to coherent, yet polarized, worldviews that define and reinforce group boundaries. In this work we propose and study a multi-topic opinion dynamics model that captures these complex inter-dependencies. Each issue is represented as a separate dimension in a shared opinion space, allowing us to model not just attitudes toward individual topics, but also the structure of ideological alignment across them. A key feature of the model is the inclusion of topic correlation, which enables us to explore how ideologically polarized states emerge when opinions on one issue influence attitudes on others. Additionally, the model incorporates homophily —a well-documented social mechanism whereby individuals are more likely to engage with and be influenced by others who are similar to themselves. We analyze the model’s asymptotic behavior by identifying and characterizing its most relevant fixed points, supported by both theoretical insights and numerical simulations. We then investigate how the multi-dimensional structure of opinion space shapes the emergence and stability of polarized states. Finally, we apply the model to empirical data from the American National Election Studies (ANES), interpreting observed opinion patterns through the lens of our framework and situating them within its parameter space.

\end{abstract}

\newpage

\section{Introduction}

Political polarization has become increasingly evident in contemporary societies~\cite{Pew_Research, Polarization_in_America}, observed both offline~\cite{Partisans_without_constraint,Myths_and_realities} and in online social media~\cite{Exposure_opposing_views,Political_participatory_media}, leading to the adoption of ever more extreme positions in public discourse~\cite{Polarization_Loop}. This phenomenon is not limited to political elites; rather, it extends to the general population~\cite{Elite_polarization}, shaping the attitudes of both leaders and ordinary citizens. As polarization deepens, it reshapes public debate, exacerbates ideological divisions, and reduces the space for compromise~\cite{Stop_talking_politics}.

One of the most concerning consequences of political polarization is its impact on democratic governance~\cite{United_we_stand}. As ideological divisions grow sharper, governments shift between radically opposed policy agendas, often reversing the work of their predecessors~\cite{Legislative_Gridlock}. This cyclical undoing of policies weakens institutional stability, undermines long-term policymaking, and fuels public distrust in democratic processes~\cite{Global_Crisis_Democracy, Affective_polarization_and_Democratic_Backsliding}.

Polarization is not merely an isolated phenomenon that affects political issues independently. Instead, it operates within broader ideological frameworks that structure political discourse into opposing camps~\cite{Partisans_without_constraint}. These ideological divisions dictate the positions that individuals adopt across multiple issues, creating a pattern in which opinions on one topic strongly correlate with positions on others~\cite{Nature_of_belief_systems,Liberals_drink_lattes}. For instance, attitudes toward transgender rights often align with views on LGBTQ+ rights more broadly, reflecting the influence of ideological alignments rather than independent assessments of each issue~\cite{Polarization_on_trasgender_rights,Public_opinion_over_gay_rights,Anxiety_factor}.

The construction of these ideological frameworks and the resulting segmentation of the population into opposing camps have been strategically leveraged by more radical political groups~\cite{Anti_pluralist_parties,Political_Disinformation_and_Hate_Speech}. With the amplification effect of social media~\cite{Technological_Capacity,Dynamics_of_collective_attention}, these groups position themselves as democratic alternatives while advancing increasingly extreme agendas. In various countries, such movements have successfully mobilized support and gained political power, further intensifying polarization and challenging democratic institutions~\cite{Global_Crisis_Democracy,United_we_stand}.

Much of the existing research on modeling polarization focuses on processes involving a single issue \cite{Voter_classic,Deffuant_classic,DeGroot_classic,Argument_1D,Baumann_1D,Success_driven_opinion,Social_phyisics_collective}. While such studies offer valuable insights, real-world political discourse rarely centers on isolated topics. Instead, it typically spans multiple, interconnected issues, with individuals forming opinions that are influenced by broader ideological frameworks. This multidimensional nature of polarization underscores the need for models that go beyond single-issue dynamics to capture the complex structure of ideological divisions.

A seminal contribution to multidimensional modeling is the Axelrod’s model of cultural dissemination~\cite{Axelrod}, which demonstrated how clusters of like-minded agents can stably coexist depending on the number of topics and features per topic. In Axelrod’s framework, the probability of interaction between agents increases with their similarity—a mechanism later formalized as homophily, the tendency of individuals to engage more with those who share similar views. This concept underlies many subsequent agent-based models, including the one proposed by Baumann et al.~\cite{Baumann_2D}, where agents hold opinions in a multidimensional topic space and form dynamic social networks, creating links preferentially with similar others. A key feature of their model is the presence of topic correlation, allowing opinions on one issue to influence others and ultimately leading to emergent ideological polarization. Topic correlation also plays a central role in the model developed by Chen et al.~\cite{Derived_Topics}, where agents begin by discussing a central issue and progressively introduce subtopics derived from it, thereby increasing the dimensionality of the opinion space over time. Similarly, in~\cite{Argument_communication_Ideology}, Banish et al. model multiple topics with their correlations defined by shared underlying arguments; the adoption of these arguments not only shapes individual opinions but also drives agents toward ideological alignment.

Other multidimensional approaches emphasize the role of social relationships. For instance, models grounded in structural balance theory~\cite{Weighted_Balance_Model, Agent_Based_Model} treat the patterns of agreement and disagreement across topics as key drivers of polarization. Agents are more likely to reinforce their agreement or disagreement depending on the coincidences in their opinions of the topics. Other models introduce special types of agents to simulate external influences. In~\cite{Influencers_and_media}, the inclusion of media actors and influencers—agents with broader reach or greater persuasive power—illustrates how mass communication channels shape polarization dynamics. In~\cite{Friedkin_Johnsen_multidimensional}, the focus is on stubborn agents, whose opinions remain fixed regardless of social interaction, and how their presence affects long-term polarization. Both these works propose potential strategies for depolarization. This theme is further explored in~\cite{Explosive_depolarization, Social_network_heterogeinity}, where the authors propose the social compass model—a polar representation of agent opinions in which the radius reflects opinion strength and the angle represents ideological direction—offering insights into the mechanisms that may lead to sudden and collective shifts toward depolarization.

Following the approach of Pérez-Martínez et al.\cite{Pol_opinions_2023}, we implemented the dynamical model proposed by Baumann et al.\cite{Baumann_2D} on a static network. In this setting homophily is incorporated into the interaction weightss: each agent assigns a weight to their neighbors’ opinions based on their degree of similarity, effectively modulating influence according to homophily. We extend the work of Pérez-Martínez et al.~\cite{Pol_opinions_2023} by generalizing their model to a multidimensional opinion space, where each dimension corresponds to a distinct political or social issue. This framework allows us to examine how polarization arises and evolves when multiple, interrelated topics are considered simultaneously. By capturing the interactions between issues, our model offers a richer perspective on how ideological alignments structure and influence opinion dynamics.

Centered on the dynamics of polarization, our agent-based model operates on complex networks and is governed by three key parameters. The first captures social influence, modeling how agents adjust their opinions in response to those of their neighbors. The second incorporates homophily, as previously described, reflecting the tendency of individuals to be more influenced by similar others. The third introduces topic correlation, measuring the interdependence of opinions across different issues. Together, these parameters enable us to explore the emergence of ideological polarization and assess how specific structural features may amplify or suppress extreme divides.

In the remainder of this paper, we begin by describing the final states of the model and the metrics used to categorize them. Next we present a theoretical analysis in which we identify all fixed points and perform a stability analysis of the most representative polarized states. We derive the conditions under which these states are stable and show that their stability depends on critical values of homophily, which vary with the type of polarization. Next, we conduct a detailed exploration of the model across the parameter space, identifying distinct regions associated with consensus, one-dimensional polarization, two-dimensional uncorrelated polarization and ideological polarization. We then study the effect of having multiple dimensions in the polarization of the agents, comparing the distributions of agents polarized in both topics versus the distributions of agents polarized in only one topic. Finally, we analyze real-world opinion data from the American National Election Survey (ANES)~\cite{ANES_2020}. By combining responses from pairs of political questions, we construct two-dimensional opinion distributions and compare them to the model’s output to assess whether the polarized states predicted by the model are reflected in empirical data.

\section{The Model}

We analyze a system composed of $N$ interacting agents, where each agent updates its opinions by considering the opinions of its neighbors within the network. The set of neighbors of each node is static and determined by the entries of the unweighted undirected adjancency matrix ${\bf A}$,  with $A_{ij} = 1$ indicating a link between agents $i$ and $j$ and $A_{ij} = 0$ otherwise. Conversely, the influence of agents on others' opinion is encoded in a temporal, weighted and directed network ${\bf w}$, where $w_{ij}$ determines how the opinion of node $i$ changes following the interaction with agent $j$.


Each agent’s opinions are represented by a vector $\mathbf{x}_i\in$  $\mathbb{R}^{m}$ in the multidimensional space of topics. The dimensionality $m$ of this space corresponds to the number of distinct topics under consideration, with each component of $\mathbf{x}_i$ reflecting the agent’s opinion on a specific topic. The sign of each component indicates the agent’s stance on the topic (positive for supporting, negative for opposing), while the magnitude represents the conviction or degree of radicalization. Thus, $\mathbf{x}_i$ encapsulates both the direction and intensity of an agent's beliefs, enabling a nuanced representation of opinion dynamics across multiple interconnected topics.

\subsection{Dynamical equations}

For the opinion formation process, we followed the approach developed in \cite{Baumann_2D, Pol_opinions_2023}, which assumes that agents' opinions are influenced by their neighbors, and they change over time according to these basic principles:
\begin{itemize}
    \item (\textit{i}) agents lose memory of their opinions with time,
    \item (\textit{ii}) they tend to align their opinions with their neighbors, and
    \item (\textit{iii}) they give more relevance to like-minded neighbors.
\end{itemize}
Additionally, in this work, we also consider agents discussing two topics simultaneously, so we add a fourth principle:
\begin{itemize}
    \item (\textit{iv}) topics can be correlated, and the neighbors' opinions in one topic affect the change in the other topic for correlated issues.
\end{itemize}

Moreover, we assume that the changes in the opinions occur faster than those in the network, so the network structure remains unchanged over time.
Taking these mechanisms into account, we can write the dynamic equations for the time evolution of the opinions of agent \textit{i} in a two-dimensional ($m=2$) topic space as:
{\large
\begin{align}
    \dot x_i^{(1)} &= - x_i^{(1)} + K \sum_j A_{ij} w_{ij} \tanh\left( [x_j^{(1)} + \cos(\delta) x_j^{(2)}] \right)
    \label{eq:one} \\
    \dot x_i^{(2)} &= - x_i^{(2)} + K \sum_j A_{ij} w_{ij} \tanh\left( [x_j^{(2)} + \cos(\delta) x_j^{(1)}] \right) 
    \label{eq:two}
\end{align}
}
where $x_i^{(1,2)}$ is the opinion of agent \textit{i} about topic $(1,2)$, \textit{K} is the \textit{social interaction strength}, which determines the strength that neighbors' opinions have over the agent's $i$ opinions, $cos(\delta)$ is the correlation between the topics. The weights $w_{ij}$ are computed as follows:
{\large
\begin{align}
    \omega_{ij} &= \frac{(d_{i,j}+\epsilon)^{-\beta}}{\sum_{l=1}^N A_{il} (d_{i,l}+\epsilon)^{-\beta}} = \frac{(||\textbf{x}_i-\textbf{x}_j||+\epsilon)^{-\beta}}{\sum_{l=1}^N A_{il} (||\textbf{x}_i-\textbf{x}_l||+\epsilon)^{-\beta}}
    \label{eq:three}
\end{align}
}
where $\beta$ is the \textit{homophily parameter} (the higher it is, the more influential like-minded neighbors become), and $\epsilon$ is included as a small noise, taken to be $\epsilon = 0.002 \, K$ following what was done in~\cite{Pol_opinions_2023}.  Note that, as stated above, the values of the weights can change over time, fulfilling the normalization condition: $\sum_j w_{ij} = 1 $. Notice also that the weights might not be symmetric,
since $w_{ij} \neq w_{ji}$.

It is important to note that the correlation between both issues makes the space non-orthogonal, such that the euclidean distance is computed by:
\begin{align}
    || \mathbf{x}|| = ||(x^{(1)},x^{(2)})|| = \sqrt{(x^{(1)})^2 + (x^{(2)})^2 + 2 \cos(\delta) x^{(1)} x^{(2)}}
    \label{eq:four}
\end{align}

\begin{figure}[H]
    \centering
    \includegraphics[width=0.8\textwidth]{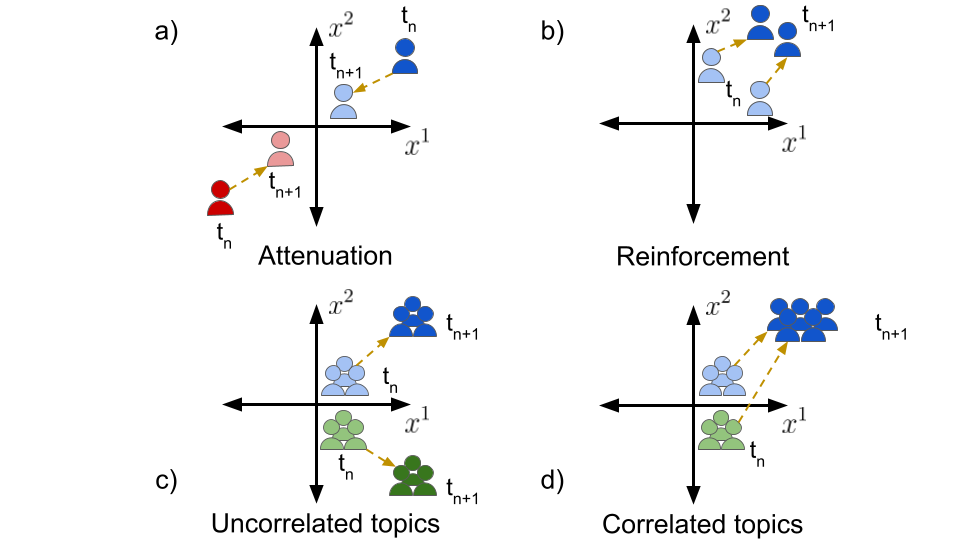}
    \caption{  \textbf{Illustrative scheme of the interactions between agents.}   (a) \textbf{Attenuation:} Agents with opposing stances decrease their convictions after interacting, drawing them closer. In this case the attenuation is applied in both axes, but it could be applied in only one. (b) \textbf{Reinforcement:} Agents share similar stances and when they interact they increase their convictions. (c) \textbf{Uncorrelated topics:} The lack of correlation let's agents freely adopt any stance, and each community has a stronger reinforcement effect within itself than attenuation with the rest, polarizing within each respective quadrant. (d) \textbf{Correlated topics:} Positive correlation favors having the same stance in each topic, which draws agents to the favored ideological positions. The disfavored quadrants become empty while agents align with the correlation.}
    \label{fig:Interaccion_agentes}
\end{figure}

As previously noted, agents update their opinions at each time step according to Eqs.~(\ref{eq:one}) and (\ref{eq:two}). Pairwise interactions between agents exhibit two distinct behaviors, emerging from the group polarization mechanism: \textit{attenuation} and \textit{reinforcement}, which are illustrated in Fig.~\ref{fig:Interaccion_agentes}. In this figure, agents are colored by their convictions, with darker shades indicating stronger convictions and different colors representing opposing stances. The interactions portrayed are:

\begin{itemize}
    \item \textbf{Attenuation} (Fig.~\ref{fig:Interaccion_agentes}a): Agents with strong initial convictions ($|x^{(1)}_i|\gg1$) and opposing stances ($sgn(x^{(1)}_i) = - sgn(x^{(1)}_j)$) reduce each other’s conviction levels through interaction, converging toward intermediate positions in the topic space. Here, two agents interact—one holding positive opinions on both topics, the other negative. When they interact, each agent updates their opinions according to Eqs. (\ref{eq:one}) and (\ref{eq:two}). The agent with positive opinions receives a negative contribution from the interaction term (due to the other agent’s opposing stance), which shifts its opinions toward more negative values. Conversely, the agent with negative opinions is influenced in the opposite direction: the interaction term becomes positive, nudging its opinions toward the positive side. As a result, both agents move toward more moderate positions in the opinion space.
    \item \textbf{Reinforcement} (Fig.~\ref{fig:Interaccion_agentes}b): Agents with similar stances ($sgn(x^{(1)}_i) = sgn(x^{(1)}_j)$) amplify each other’s convictions, resulting in more extreme opinions after interaction. If both agents have positive stances, the hyperbolic tangent term in the dynamic equations becomes positive, pushing their stances for even further positive values.
    \item \textbf{Effect of correlation in topics:} (Fig.~\ref{fig:Interaccion_agentes}c and \ref{fig:Interaccion_agentes}d): As correlation between topics increases, the effects previously described become asymmetric and agents holding stances disfavored by the topic correlation exert weaker influence on those aligned with the correlation. Each group of agents in the figure represents a tightly connected community with weak ties to others due to homophily. In the uncorrelated case each community ends up reinforcing itself and polarizing within their respective quadrants. But as correlation grows, agents with favored opinions increasingly pull others away from less aligned positions.
\end{itemize}

\subsection{Simulations setup}

To characterize the model, we analyze a system of $N = 10000$ agents arranged in an Erdös-Rényi network with a mean degree of $\langle k \rangle = 10$.
The system of $2N$ coupled equations is solved using an explicit fourth-order Runge-Kutta method with a time step of $dt = 0.1$. The initial conditions are set by uniformly distributing each agent’s opinion in both topics within the range $[-K, K]$. The system is then allowed to evolve until it reaches equilibrium, defined as the point where agents' opinions on average no longer change over time.

The free parameters of the model are: 
\begin{itemize}
    \item the homophily between agents $\beta$,
    \item the social interaction strength  $K$
    \item the correlation between topics $\cos(\delta)$.
\end{itemize}

A total of $N_{ens} = 100$ simulations of the system were performed for each combination of the parameters $(\beta, K)$ and $(\beta, \cos(\delta))$, while keeping the third parameter fixed in each case.

\subsection{Stationary and quasi-stationary states}
\label{section:classification}

\begin{figure}[t!]
    \centering
    \includegraphics[width=0.9\textwidth]{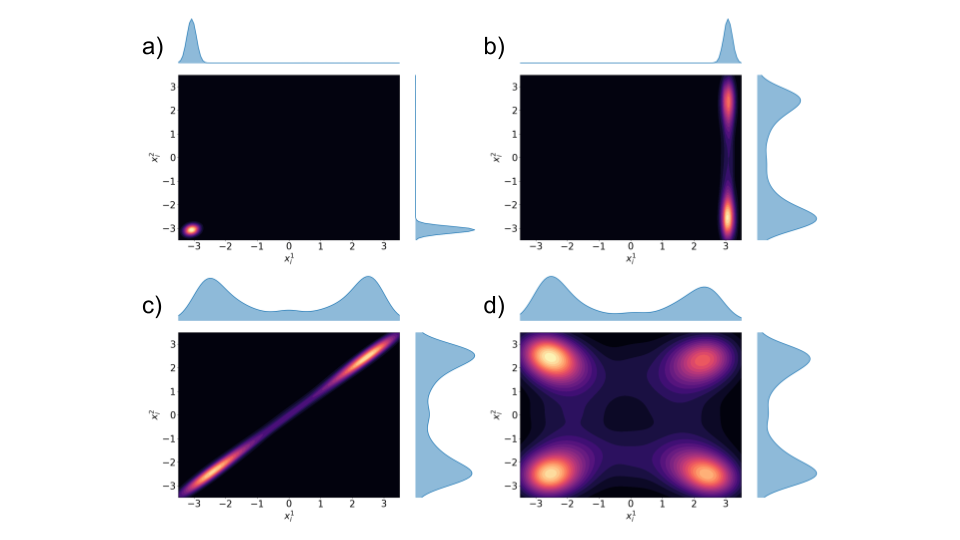}
    \caption{  \textbf{Examples of stationary and quasi-stationary states.}  The system was observed to reach four distinct macro-states, characterized by the distribution of agents in the topic space: (a) Consensus \textit{(Stationary)}, (b) 1D Polarization \textit{(Quasi-stationary)} (c) Ideological Polarization \textit{(Stationary)}, (d) Uncorrelated Polarization \textit{(Stationary)}. The model is able to reproduce states of consensus, polarization in one topic and in both topics.}
    \label{fig:Estados_finales}
\end{figure}

Depending on the combination of parameters, the system was observed to reach four distinct macro-states, characterized by the distribution of agents in the topic space as well as their distribution within each topic separately. These macro-states are illustrated in Fig.(\ref{fig:Estados_finales}).

The system reaches a state of consensus when all agents converge to a single point in the topic space. Consensus can be further classified as either neutral or radicalized, depending on whether agents hold null opinions on both topics or not, respectively. The neutral state typically emerges for low values of $K$, while the radicalized state occurs for high $K$ and low $\beta$. An example of a radicalized consensus state is shown in Fig.(\ref{fig:Estados_finales}a).

Unlike consensus states, polarized states can take on various forms. These macro-states are characterized by the division of the agent population into two or more opposing groups in the topic space, each holding divergent opinions on the topics.

One type of polarization is one-dimensional polarization, where agents polarize along one axis while reaching consensus on the other. This pattern typically arises under moderate homophily and uncorrelated topics, as shown in Fig.(\ref{fig:Estados_finales}b). It is important to note that these states are generally not truly stationary but quasi-stationary, with lifespans long enough to be treated as effectively stationary for our analysis.

When polarization occurs such that the opinion in one topic directly correlates with the opinion in the second topic, the system reaches a state of ideological polarization. In this case, agents are distributed along a diagonal in the topic space. Since we only consider positive correlation between topics, this means that agents align along the diagonal where their opinions in both topics coincide, as shown in Fig.(\ref{fig:Estados_finales}c).

Finally, polarization can occur simultaneously in both topics causing agents to divide into four distinct stances, which occupy the four corners of the topic space. Whether agents adopt only these four stances or distribute themselves across the entire topic space, we refer to these states as uncorrelated polarization. An example of such states is shown in Fig.(\ref{fig:Estados_finales}d).

To classify each final state, we used statistical measures of the opinion distribution across both topics. Polarization was identified using the opinion variances $(\sigma_1, \sigma_2)$: high variance in both topics indicates two-dimensional (2D) polarization, high variance in one topic indicates one-dimensional (1D) polarization, and low variance in both signifies consensus. Within 2D polarization, the covariance $(\sigma_{12})$ distinguishes uncorrelated polarization (low covariance) from ideological polarization (high covariance). For consensus states, we differentiated neutral from radicalized consensus using the mean conviction of agents, i.e., the mean modulus of opinions $(\langle |x_i^v| \rangle)$. The threshold values used to classify each final state were obtained through manual inspection of the data; detailed criteria are provided in the Appendix~(\ref{section: Clasificacion}) .

\section{Results}

\subsection{Analytical results}

As we mentioned in the previous section there are different equilibrium states that emerge as solutions of the system depending on the parameter combination and initial conditions. Here we study the stability of the fixed points for the cases of consensus and two-dimensional polarization.

\subsubsection{Stability of the consensus states}

First, everyone can agree on the same equilibrium opinion $\mathbf{x}^* = (x^{*(1)},x^{*(2)})$, corresponding to a radicalized state. The stationary opinion values are given by the system of equations:
\begin{align}
    x^{*(1)} &= K \tanh\left( x^{*(1)} + x^{*(2)} \cos\delta \right)\; , \\
    x^{*(2)} &= K \tanh\left( x^{*(2)} + x^{*(1)} \cos\delta \right) \; .
\end{align}

\noindent The actual values for the equilibrium opinions depend on $K$ and $\delta$. If $K$ is small enough, we only find consensus as a solution, $x^{*(1)}=x^{*(2)} = 0$. If $K$ is large enough, then there are four possible stable solutions depending on the sign of each component: either $\text{sign}(x^{*(1)}) = \text{sign}(x^{*(2)})$, or $\text{sign}(x^{*(1)}) =- \text{sign}(x^{*(2)})$. Nevertheless, in both cases opinions satisfy $|x^{*(1)}| = |x^{*(2)}|$; in the former case the fixed point fulfills $x^{*(1)} = x^{*(2)} = K\tanh\left[x^{*(1)}(1+\cos\delta)\right]$, so it appears under the condition $K(1+\cos\delta) \geq 1$. The point becomes stable if $K(1+\cos\delta)<\cosh^2\left[x^{*(1)}(1+\cos\delta)\right]$, which is always true in its range of existence. In the latter case, the fixed point follows $x^{*(1)} = -x^{*(2)} = K\tanh\left[x^{*(1)}(1-\cos\delta)\right]$, and exists if $K(1-\cos\delta) \geq 1$. The stability condition reads $K(1+\cos\delta)<\cosh^2\left[x^{*(1)}(1-\cos\delta)\right]$, so that it becomes stable provided $\cos\delta$ is small enough. We refer the reader to the Appendix~(\ref{section: Analisis radicalizacion}) for a thorough analysis of the system's stability.

\subsubsection{Stability of the two-dimensional polarization states}

Despite these radicalized consensus states we can also find polarization, as defined in section \ref{section:classification}. In general, the population of $N$ agents can split into $n$ distinct groups occupied by $\{N_1,...,N_n\}$ agents holding opinions $\{\mathbf{x}_1,...,\mathbf{x}_n\}$. Assuming a fully connected network, each equilibrium opinion $\mathbf{x}_\xi$ must fulfill the system of equations:
\begin{align}
    x_\xi^{(1)} &= K \sum_{\alpha=1}^{n} N_\alpha \omega_{\xi \alpha} \tanh(x_\alpha^{(1)} + x_\alpha^{(2)} \cos\delta)\; , \\
    x_\xi^{(2)} &= K \sum_{\alpha=1}^{n} N_\alpha \omega_{\xi \alpha} \tanh(x_\alpha^{(2)} + x_\alpha^{(1)} \cos\delta) \; ,
\end{align}

The two most relevant solutions to this system of equations are the cases in which: (\textit{i}) the population becomes split in two groups holding opinions $\{\mathbf{x}_{++},\mathbf{x}_{--}\}$, where the subindex denotes the sign of the opinion of the corresponding component (\textit{i.e.} $x_{++}^{(1)}, x_{++}^{(2)}>0$, and $x_{--}^{(1)}, x_{--}^{(2)}<0$), resulting in what we call \textit{ideological polarization}, and (\textit{ii}) the population splits into four groups holding opinions $\{\mathbf{x}_{++},\mathbf{x}_{+-},\mathbf{x}_{-+},\mathbf{x}_{--}\}$, producing a state which we call \textit{uncorrelated polarization}.

Now, to check the stability of such equilibrium situations, we introduce a perturbation $\boldsymbol{\lambda} = \lambda (\cos\varphi, \sin\varphi)$ with $\varphi\in [0,2\pi)$ in an agent $i$ holding initially an opinion $\mathbf{x}_i = \mathbf{x}_\xi$. Then, from the dynamical equations for the perturbed agent we can find the following equation for the module of the perturbation:
\begin{equation}
    \dot{\lambda} = \dot{x}_i^{(1)} \cos\varphi + \dot{x}_i^{(2)} \sin \varphi \; .
\end{equation}

\noindent Assuming $\lambda\ll 1$, we can write approximate equations for $\dot{x}_i^{(1)}$, $\dot{x}_i^{(2)}$ as:
\begin{align}
    \dot{x}_i^{(1)} &\simeq -\lambda \cos \varphi + K\lambda \beta \frac{B^{(1)}C - A^{(1)}D^{(1)}}{C^2}\; , \\
    \dot{x}_i^{(2)} &\simeq -\lambda \sin \varphi + K\lambda \beta \frac{B^{(2)}C - A^{(2)}D^{(2)}}{C^2}\; ,
\end{align}

\noindent where
\begin{align*}
    A^{(1,2)} &= \sum_{\alpha=1}^n N_\alpha f_{\xi\alpha}^\beta \tanh\left(x_\alpha^{(1,2)} + x_\alpha^{(2,1)}\cos\delta\right) \; , \\
    B^{(1,2)} &=\sum_{\alpha \neq \xi} N_\alpha f_{\xi\alpha}^{\beta-1}g_{\xi\alpha} \tanh\left(x_\alpha^{(1,2)} + x_\alpha^{(2,1)}\cos\delta\right) \; , \\
    C &= \sum_{\alpha=1}^n N_\alpha f_{\xi\alpha}^\beta\; , \\
    D &= \sum_{\xi\neq\alpha}N_\alpha f_{\xi\alpha}^{\beta-1}g_{\xi\alpha} \; ,
\end{align*}

\noindent and:
\begin{align*}
    f_{\xi\alpha} &= \frac{\epsilon}{d_{\xi\alpha}+\epsilon} \; , \\
    g_{\xi\alpha} &= \frac{\sqrt{1+2\sin\varphi\cos\varphi \cos\delta}(d_{\xi\alpha}+\epsilon) +\epsilon \frac{ (x_\alpha^{(1)}-x_\xi^{(1)})(\cos\varphi + \sin\varphi\cos\delta) + (x_\alpha^{(2)}-x_\xi^{(2)})(\sin\varphi + \cos\varphi\cos\delta) }{d_{\xi\alpha}}}{(d_{\xi\alpha}+\epsilon)^2} \; .
\end{align*}

\noindent By computing the value of $\dot\lambda$, we can check if the system is stable ($\dot\lambda<0$) or unstable ($\dot\lambda>0$) under the proposed perturbation. From the previous equations we can finally obtain the general condition:
\begin{equation}
    K\beta \frac{(B^{(1)}C-A^{(1)}D)\cos\varphi + (B^{(2)}C-A^{(2)}D)\sin\varphi}{C^2} < 1 \; .
\end{equation}

\noindent If this condition applies for all opinion clusters and all possible perturbation directions (\textit{i.e.} all possible $\varphi$), then we conclude that the system is stable. This condition can be solved numerically for both systems, ideological and uncorrelated polarization (for a thorough analysis of the system's behavior, see Appendix~(\ref{section: Analisis polarizacion})). Assuming an homogeneous partition into groups of the same size, $N_\alpha\simeq N/n, \, \alpha=1,...,n$, and a high enough coupling $K$, we find that the stability condition for ideological polarization is $\beta_c^{\text{ideol.}}=1$, and for uncorrelated polarization, $\beta_c^{\text{uncor}} >1$. In particular, for the case $\cos\delta=0$, we find that $\beta_c^{\text{uncor}}\simeq 1.145$, and it increases monotonously for higher values of $\cos\delta$. Therefore, we find that the existence of additional ideologies, understood as the presence of more opinion clusters, makes polarization more difficult. As topics become more correlated, these opinion clusters will tend to disappear and ideological polarization will become the only stable situation.

\subsection{Numerical results}

\begin{figure}[H]
    \centering
     \includegraphics[width=1\textwidth]{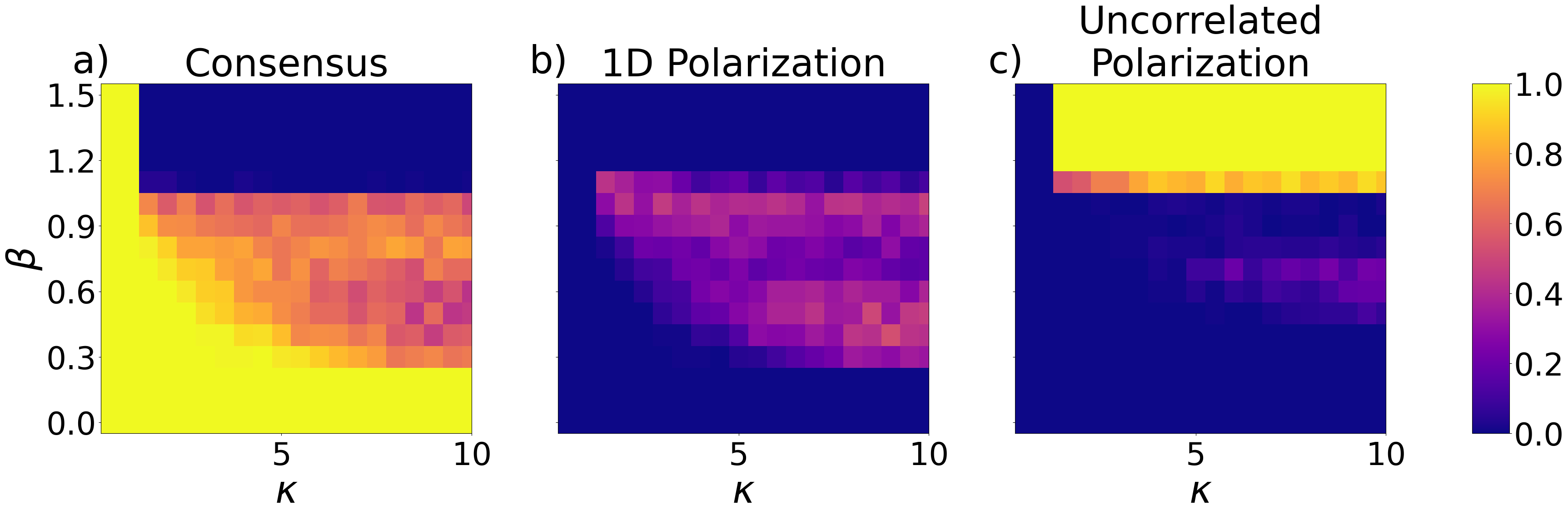}
    \caption{ \textbf{Fraction of the of the total final states in the parameter space for:} (a) Consensus, (b) One dimensional polarization and (c) Uncorrelated polarization. Results correspond to an ER network of $\langle k \rangle=10$ and $N=10000$ nodes. Throughout the figure, $\cos\delta=0$. We characterize the model using density maps corresponding to each of the final states. States of ideological polarization do not emerge, as the correlation between topics is zero. }
    \label{fig:Esp_BK}
\end{figure}

We analyze and classify the final states of the model in the $(\beta - K)$ space. Setting $\cos(\delta) = 0$, we simulated the model for $\beta \in [0,1.5]$ and $K \in [0,10]$. According to Section~(\ref{section:classification}), final states are categorized as consensus, one-dimensional polarization, or uncorrelated polarization using the variance, covariance, and mean conviction of each distribution. Since topics are uncorrelated, no ideological polarization states emerge. After classifying all states in the ensemble, we compute the fraction of each state at every point in the parameter space as the number of simulations in that state divided by the total simulations at that point, $N_{ens}$. Fig.(\ref{fig:Esp_BK}) shows the state density maps constructed in the parameter space.  

Figs.(\ref{fig:Esp_BK}a) and (\ref{fig:Esp_BK}c) reveal two distinct regions where the system reaches only either a consensus, when $\beta$ is low enough, or uncorrelated polarization states, when $\beta>\beta_c^{\text{uncor}}$, thus validating our previous mathematical analysis. 
Between these regions, for $\beta < 1$ and $K > 1$, a mixed regime appears, where the system can mostly converge to either consensus or one-dimensional polarization. This region shows different ratios of each state that varies with the parameters $\beta$ and $K$. The presence of equilibrium one-dimensional polarized states for $\beta<1$, where they shouldn't be stable, is due to the presence of a networked contact structure that imposes a limited information horizon to the agents. The network modulates the exposure of agents to cross-cutting relationships and hampers depolarization processes. The resulting states are not static, but rather consist on a dynamic quasi-equilibrium that eventually depolarizes, albeit with a very long lifespan that renders them effectively stable. This effect is also present in the one-dimensional version model (see \cite{Pol_opinions_2023} for a more thorough analysis). Note that one-dimensional polarization is still a stable solution of the system for $\beta>1$, but it is hardly reached when uncorrelated polarization is also stable, as it becomes the favored final states thanks to the random initial conditions.

We conclude that the introduction of two different topics in the opinion space does not alter the qualitative behavior of the system significantly, the main difference being that polarization for $\beta>\beta_c^{\text{uncor}}$ brings the presence of four different opinion combinations in the form of uncorrelated polarization, as opposed to two. Apart from that, we see that most equilibrium states show consensus either on one or both axis at the same time. Therefore, most of the time the equilibrium states resemble those of the original one-dimensional model. Regardless, for high values of $K$ we find some bidimensional polarized states, unique to the bidimensional model.

\begin{figure}[H]
   \centering
     \includegraphics[width=0.7\textwidth]{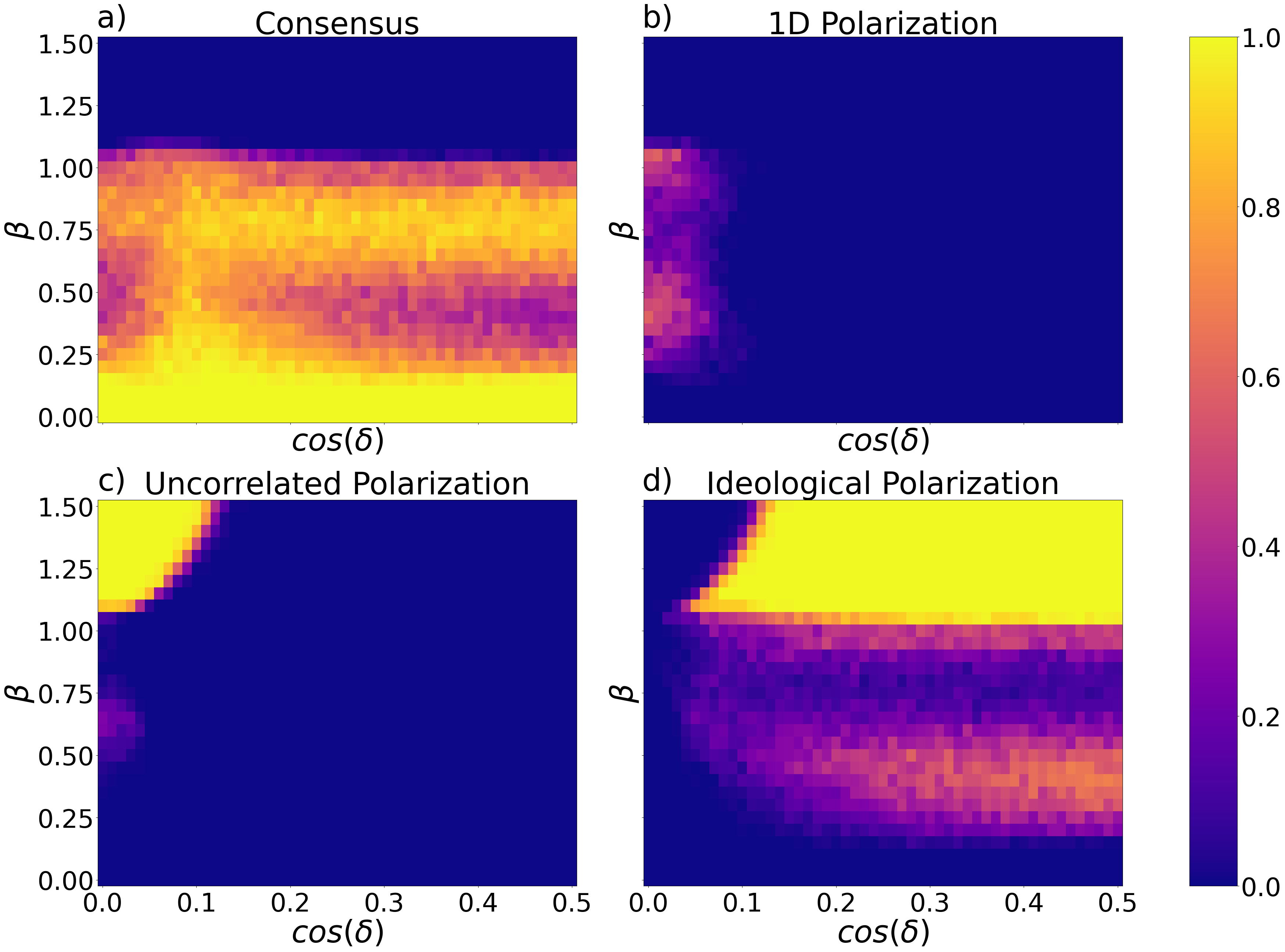}
    \caption{  \textbf{Fraction of the of the total final states in the parameter space for:}  (a) Consensus, (b) One dimensional polarization, (c) Uncorrelated polarization and (d) Ideological polarization. Results correspond to an ER network of $\langle k \rangle=10$ and $N=10000$ nodes. Throughout the figure, $K=10$. The model is characterized through the density maps corresponding to each of the final states.}
    \label{fig:Esp_BCD}
\end{figure}

To investigate how the correlation between topics influences the equilibrium states, we simulate the model within the intervals $[0, 1.5]$ for $\beta$ and $[0, 0.5]$ for $\cos(\delta)$, using $K = 10$. We classify the final states as in Fig.~(\ref{fig:Esp_BK}) and construct the density maps shown in Fig.(\ref{fig:Esp_BCD}).

We observe that a clear transition occurs again at $\beta\simeq1$ between a region dominated by consensus states for $\beta<1$, and other dominated by polarization for $\beta>1$. The type of polarization depends on $\cos\delta$: low topic correlation leads to uncorrelated polarization, while high correlation results in ideological polarization. It is important to note that the transition from consensus to polarization occurs around $\beta\simeq 1.1$ for $\cos\delta=0$ because it is dominated by uncorrelated polarization, but it quickly decays towards $\beta=1$ for increasing values of $\cos\delta$ because polarized states become increasingly correlated, and thus, stable for lower values of $\beta$. Regarding the transition between correlated and uncorrelated polarized states for $\beta>1$, it is clear that higher values of $\beta$ allow for uncorrelated polarization to be maintained even for correlated topics, as homophily decreases the influence that individuals from different opinion clusters exert on each other.

In the intermediate region for $\beta<1$, we find again a mixture of consensus and polarization states. However, one-dimensional polarization and uncorrelated polarization quickly disappear in favor of ideological polarization. Therefore, even for very weak correlation between topics, the system can quickly become sorted between only two opposed groups.

\subsection{The one dimensional projection: comparison with the 1D model}

In a multidimensional opinion space, agents who polarize on one topic may still agree on others. In such cases, ideological homophily can make them more receptive to each other’s views, increasing the likelihood of reaching agreement on additional topics. This mechanism can lead to ideological sorting, even when the issues themselves are unrelated. This process can affect the polarization obtained if we compare the polarization distributions of our two-dimensional model with the one-dimensional model, in which agents don't have a secondary topic in which they could agree. We analyze then how 1D polarization states behave in the multidimensional model and assess whether the presence of a second topic influences the degree of polarization. We also examine whether this difference diminishes as topic correlation increases, causing opinions to align along a single ideological axis.

\begin{figure}[H]
    \centering
     \includegraphics[width=0.7\textwidth]{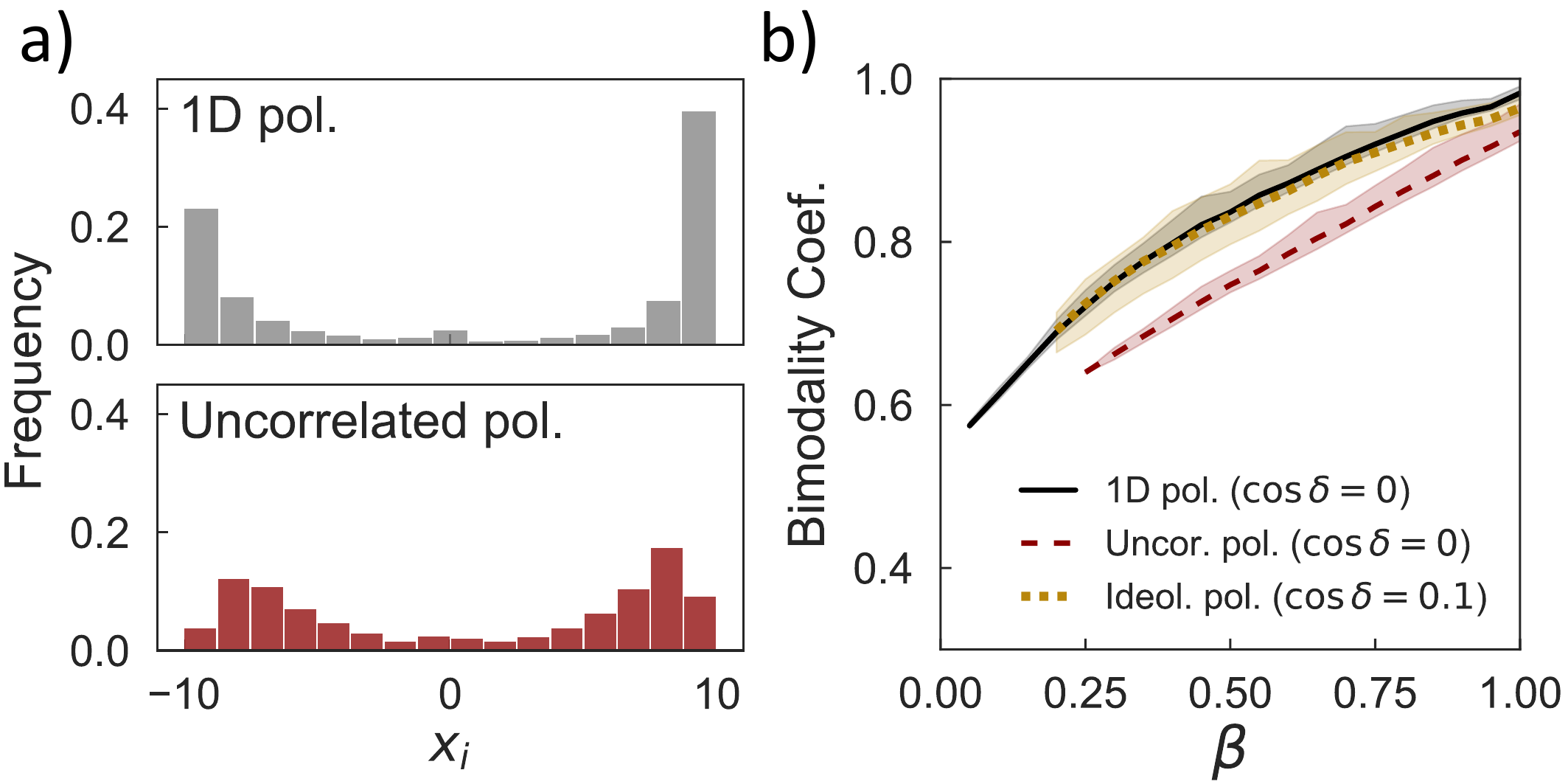}
    \caption{\textbf{Comparison of the opinion distributions between one-dimensional and two-dimensional polarization:} (a) Opinion distributions projected on a single axis for (top) one-dimensional polarization and (bottom) bidimensional, uncorrelated polarization, taking $\cos\delta=0$, $K=10$ and $\beta=0.6$. (b) Bimodality coefficient of the polarized opinion distributions projected on a single axis as a function of $\beta$, for one-dimensional polarization (black, solid line) and uncorrelated polarization (red, dashed line) in the case of uncorrelated topics ($\cos\delta=0$), and ideological polarization (yellow, pointed line) for correlated topics ($\cos\delta=0.1$). Shadowed regions correspond to $95\%$ confidence intervals, obtained from 100 independent configurations for each parameter combination of $\beta$ and $\cos\delta$. The opinion distributions of the uncorrelated polarization states projected into one dimension are less polarized than the one-dimensional polarization states. As correlation increases the system becomes \textit{effectively} one-dimensional and the distributions become more polarized, as shown by the bimodality coefficient.}
    \label{fig:2Total}
\end{figure}

We first consider the case of two uncorrelated topics ($\cos\delta = 0$) with absence of homophily. Polarization can then emerge in two different ways: society can be polarized in only one issue while agreeing on the other, or it can become polarized in both issues at the same time. The first case is completely equivalent to the original 1D model. In Fig.(\ref{fig:2Total}a) we show the resulting opinion distributions for the same parameter combination when polarization occurs in only one dimension (1D polarization), and the opinion distribution of one of the two topics when polarization occurs in both at the same time (uncorrelated polarization), considering the other topic a secondary hidden dimension for the sake of the analysis. The different shape between both distributions suggests that the presence of a secondary polarized issue can impact the opinions about a different topic even in the absence of correlation between topics. In fact, when the population is distributed across the whole bidimensional opinion space in an uncorrelated polarization state, agents tend to adopt milder views and become less radical. 

To further verify this effect, we compute the bimodality coefficient~\cite{pfister2013bimodality} for multiple polarized opinion distributions of both kinds, one-dimensional and uncorrelated polarization, which we represent in Fig.(\ref{fig:2Total}b). It becomes clear that polarization on a secondary issue affects the opinion distributions of the main topic helping agents to adopt more moderate positions, as the bimodality coefficient becomes systematically smaller for the same parameter combination. This effect is present without considering correlation, so the existence of two topics only influences the dynamics by altering the agents' perception about their neighbors. Even if they differ in one topic, agreeing on the other makes them feel closer, and thus, they become more persuasive on the first issue.

As there is no correlation between topics, all possible opinion combinations are equally possible, so it is relatively easy to agree in one topic while disagreeing on the other. However, in the presence of correlated topics ($\cos\delta\neq 0$), it becomes difficult to hold opinions outside of the ideology (both positive or both negative at the same time), and therefore, agreeing in only one issue becomes increasingly uncommon. If $\cos\delta$ is high enough, ideological polarization becomes the only polarized solution of the system. As can be seen in Fig.(\ref{fig:2Total}b), the bimodality coefficients of such opinion distributions become equal to the 1D case, resulting in the same opinion distributions and indicating that the dynamics of the system become effectively one-dimensional, taking ideology as the new dimension in which the agent holds the same opinion on both issues. 

Our results show that an increased sorting by ideologies (which has been linked to polarization growth and political anger~\cite{iyengar2019affective,baldassarri2020culturewar}) is detrimental to the society, as individuals become increasingly detached from their cross-cutting relationships and therefore more radicalized.

\section{Survey data analyzed within the framework of the model}

In this last section, we aim to assess how well our model describes the distribution of opinions across different topics drawn from survey data. In particular, we examine whether these data correspond to categories similar to those identified in our model, and how they can be situated within the analyzed parameter space.

\subsection{Experimental data}

We used opinion data from the 2020 ANES survey~\cite{ANES_2020}, a long-standing survey conducted across the United States that covers various political topics. We extracted the distributions of responses to eleven survey questions and paired them to construct a total of 55 histograms of opinions in a 2D space (for more information on the questions considered see the Appendix~(\ref{section: Encuesta})). By using the Jensen-Shannon distance, which employs the Kullback-Leibler divergence to quantify the similarity between two distributions, we perform a quantitative analysis to categorize the data into states similar to the ones produced by our model.

The Kullback-Leibler divergence is calculated as:
\begin{align}
    KLD(P||Q) = \sum_{x \in \chi} P(x) log \left( \frac{P(x)}{Q(x)} \right)
\end{align}

With $P(x)$, $Q(x)$ two probability distributions and $\chi$ the set of discrete values. This divergence isn't symmetric with respect to the distributions considered. However, we can use it to calculate the Jensen-Shannon distance, which is a symmetric measure computed as follows:
\begin{align}
    JSD(P||Q) = \sqrt{\frac{1}{2} KLD(P||M) + \frac{1}{2} KLD(Q||M)}
\end{align}

Where $M = \frac{1}{2} (P+Q)$ is a mixture distribution of $P$ and $Q$. This Jensen-Shannon distance is zero when the distributions $P$ and $Q$ are the same and grows tending to 1 as they become increasingly different.

In the survey, we observed that a significant proportion of respondents chose not to take a stand on certain topics. In our model, this kind of inactive agents is not considered, so we exclude those individuals to perform the comparison with the Jensen-Shannon distance. Nevertheless, their inclusion would alter the computed distances similarly for all comparisons, leaving our results qualitatively equal.

By computing the Jensen-Shannon distance between the bidimensional distributions obtained from histograms of survey opinions we construct a similarity matrix. We identify distinct states within the histograms by applying a K-means clustering algorithm to the similarity matrix. Using the Silhouette Coefficient, we determine that partitioning the dataset into two clusters provide the optimal classification. The first cluster consists of radicalized consensus and unidimensional polarization states while the second contains states of uncorrelated and ideological polarization states. This partition demonstrates the presence of a clear classification of states within the data. This process is exemplified in the upper panels of Fig.(\ref{fig:Pipeline}).

\subsection{Comparison with the model}

\begin{figure}[H]
    \centering
     \includegraphics[width=0.9\textwidth]{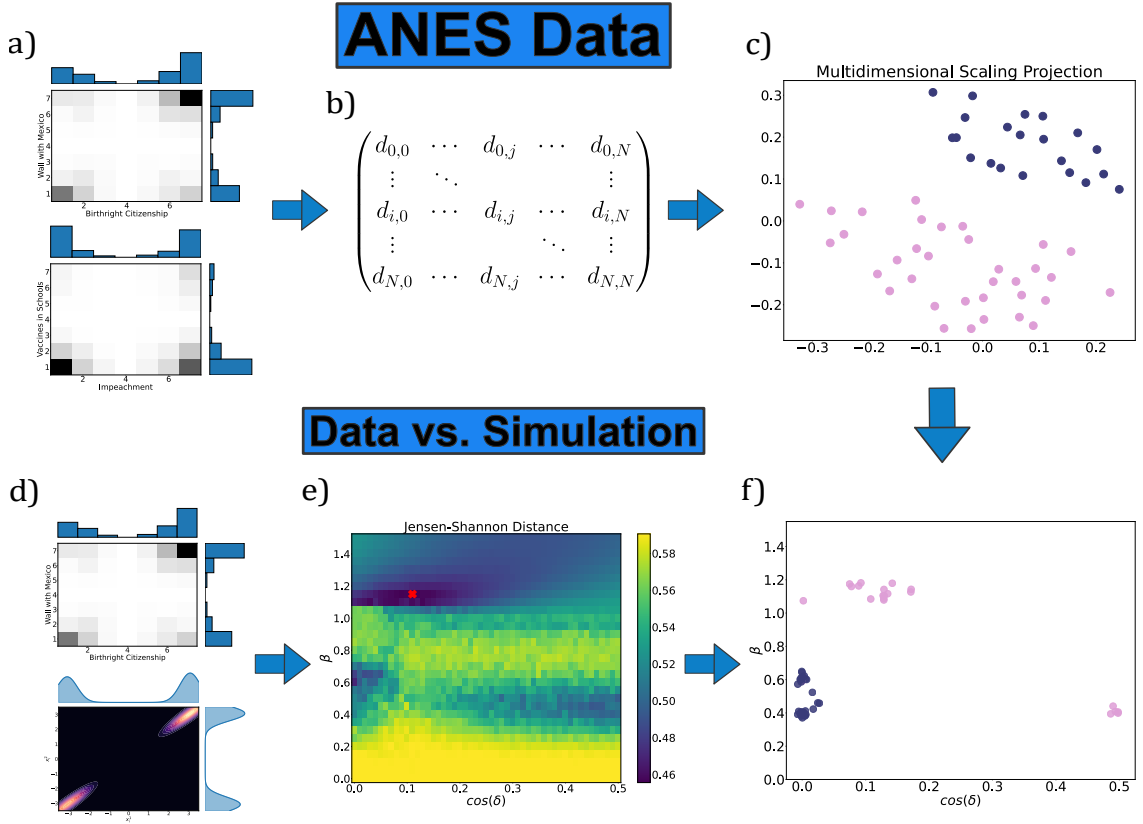}
    \caption{\textbf{Survey data classification and mapping in the parameter space $(\beta, \cos(\delta))$.} Upper panels (a,b,c): Process of clustering the survey data. This involves constructing a similarity matrix based on the Jensen-Shannon distance between the survey histograms, applying K-means clustering to the matrix, and determining the optimal partition using the Silhouette coefficient. The resulting clusters are then projected into a 2D space using a Multidimensional Scaling (MDS) method. Lower panels (d,e,f): Calculation of the Jensen-Shannon distance between the survey histograms and all simulations in the parameter space $(\beta, \cos(\delta))$, followed by identifying the parameter combination that minimizes the mean distance. The distribution of all survey histograms is then plotted in the parameter space, with clusters colored according to the upper-panel classification. The clusters obtained using the similarity of the data distributions appear separated in the parameter space, indicating that our model differentiates the data in the same way as when comparing the data to itself.}
    \label{fig:Pipeline}
\end{figure}

Our goal is to perform a quantitative mapping of the survey data onto the parameter space of the model $(\beta, \cos(\delta))$ as a way to analyze the data within the framework of our model. To perform this analysis, we generate 100 opinion distributions for each possible parameter combination $(\beta, \cos(\delta))$, producing an ensemble that includes all possible states considered in Fig.(\ref{fig:Esp_BCD}). The lower panels of Fig.(\ref{fig:Pipeline}) illustrate how we map the survey data to a point in the parameter space by comparing the opinion histograms from the survey with those generated by the simulations. 

First, we compute the Jensen-Shannon distance between the survey data and each simulation in the ensemble, and then calculate the mean distance for each point in the parameter space. The parameter combination that minimizes the mean Jensen-Shannon distance is identified as the corresponding point in the parameter space that generates the most similar opinion distributions. By repeating this process for all survey histograms, we obtain their distribution in the parameter space.

The distribution of survey histograms in the final graph of Fig.~(\ref{fig:Pipeline}) reveals that the previously identified clusters separate into two main groups. The purple dots represent the one-dimensional polarization and consensus states which are concentrated in the low-homophily, zero-correlation region, consistent with consensus and 1D polarized states. The pink dots on the other hand represent the two-dimensional polarization states that are split into two distinct regions: One with high homophily and low correlation where the transition between uncorrelated polarization and ideological polarization occurs, and the second with high correlation and low homophily where a mix of ideological polarization and consensus states coexist.

This shows a direct mapping between survey states and the model's stationary states in the $(\beta, \cos(\delta))$ parameter space, which indicates that the clusterization obtained analyzing the ANES data reflects the differences in polarization of the histograms.

\section{Conclusions}

In this work, we proposed a dynamic opinion model in a multidimensional space as a generalization of previously studied one-dimensional frameworks. We have characterized the system’s behavior as a function of three key parameters: the influence of neighboring agents, the degree of homophily, and the correlation between topics, showing that polarization can emerge not only under high levels of homophily, but also in regions with low homophily driven by the network structure.
Furthermore, our model allows for the coexistence of qualitatively different polarized and consensus states for the same parameter combination, including states of one-dimensional polarization, and bidimensional polarization, both uncorrelated and ideological. All these states reproduce qualitatively what can be observed in real-world surveys about polarized issues.

We have also shown mathematically that the stability condition for polarization across multiple topics at the same time varies significantly from the case of a single topic, highlighting the richness and complexity that arises from opinion dynamics in multidimensional spaces. More specifically, the presence of multiple opinion combinations make agents feel closer to neighbors with some cross-cutting opinions, thus making polarization more difficult and requiring higher homophily to be achieved. An increased ideological sorting, understood as a reduced number of present ideologies, progressively decreases the complexity of the system rendering the dynamics effectively one-dimensional.

We also analyzed how the dimensionality of the topic space influences polarized opinion distributions by comparing two-dimensional polarized states with their one-dimensional counterparts. Our results show that opinion distributions are sensitive to the presence of multiple topics, even when these topics are uncorrelated. However, as said before, when topic correlation is strong enough to eliminate disfavored ideologies, the system effectively reduces to a one-dimensional spectrum, with agents aligning along a single ideological axis. These findings highlight the importance of using multidimensional models to study polarization, as they capture richer and more complex dynamics that are absent in simpler, single-topic frameworks.

To explore the empirical relevance of our model, we analyzed real-world data from the American National Election Studies (ANES). We constructed bidimensional opinion distributions from the survey responses and used the Jensen–Shannon distance to categorize each possible issue combination and quantify the similarity between these empirical distributions and the equilibrium states generated by our model. This allowed us to classify observed opinion profiles according to whether polarization occurred in both ideological axes.

Polarization is a complex and multifaceted phenomenon that lacks a single, universally accepted definition~\cite{Mass_Polarization}. In this study, we focused on a specific aspect of polarization: the division of a population into two sharply opposing positions on a given issue. When extended to a multidimensional context, this phenomenon becomes even more nuanced, incorporating ideological polarization—where individuals interpret information and form opinions through the lens of a broader ideological framework. In this sense, our work is limited in scope, as it considers only two possible issues at the same time. However, it provides a useful generalization of previous models that allows for the study of multiple issues with different degrees of correlation.

Despite the uncertainty surrounding the full implications of polarization for democratic societies, it is clear that this process is actively exploited and intensified by elites for political and economic purposes. We believe that contributing to the understanding of polarization through a multidimensional opinion model—one that incorporates homophily in social interactions and correlations between discussion topics—is a meaningful step toward clarifying how and why societies become polarized.

\section*{Acknowledgments}
F.d.C. and P.B acknowledge the support of the Universidad de Buenos Aires (UBA) through Grant UBACyT, 20020220100181BA and the Agencia Nacional de Promoción de la Investigación, el Desarrollo Tecnológico y la Innovación through Grant No. PICT-2020-SERIEA-00966. H.P.M. and J.G.G. acknowledge support from Departamento de Industria e Innovaci\'on del Gobierno de Arag\'on y Fondo Social Europeo (FENOL group grant E36-23R) and Ministerio de Ciencia e Innovaci\'on (grant PID2023-147734NB-I00). D.S.-P. acknowledges financial support through grants JDC2022-048339-I and PID2021-128005NB- C21 funded by MCIN/AEI/10.13039/501100011033 and the European Union “NextGenerationEU”/PRTR”.
\newpage

\printbibliography

\newpage

\section{Appendix}

\subsection{Classification of states}
\label{section: Clasificacion}

The final states obtained from our simulations are represented by two dimensional opinion distributions. To classify each of these states into one of four categories—consensus, one-dimensional polarization, uncorrelated polarization, or ideological alignment—we employed three statistical measures: the variance of opinions along each topic $(\sigma_{xx}^2,\sigma_{yy}^2)$, the covariance of opinions between topics $(\sigma_{xy}^2)$ and the mean of the agents' convictions $(\langle |x| \rangle)$.


When analyzing the opinion distributions of the final states, it is important to take into consideration the fact that the range of opinion values varies with the parameter $K$, which represents the social interaction strength. As a result, two states that are qualitatively equally polarized may exhibit significantly different variance values solely due to differences in opinion range. To enable consistent comparison across all final states, we normalize the opinion values to lie within the interval $[-1,1]$. This is achieved by rescaling each agent’s opinion by dividing it by the value of $K$. The motivation for this normalization becomes evident when examining the dynamical equations of the model.

\begin{align*}
    \dot x_i^{(v)} &= - x_i^{(v)} + K \sum_j A_{ij} w_{ij} \tanh\left( [x_j^{(v)} + \cos(\delta) x_j^{(p)}] \right)
\end{align*}

In this case we are considering the dynamical equation for the opinion of agent \textit{i} on topic \textit{v}. If we consider a fixed point $\overrightarrow{x}^*$ such that $\left. \overrightarrow{\dot x} \right|_{\overrightarrow{x}^*} = 0$, then we can rewrite the previous equation as:

\begin{align*}
    x_i^{*(v)} &= K \sum_j A_{ij} w_{ij} \tanh\left( [x_j^{*(v)} + \cos(\delta) x_j^{(*p)}] \right)
\end{align*}

To show that no matter the solution to the equation the opinions of the agent \textit{i} will be contained in the range $[-K,K]$, we just need to take absolute value of the equation and then apply the triangle inequality.

\begin{align*}
    \left| x_i^{*(v)} \right| & = \left|K \sum_j A_{ij} w_{ij} \tanh\left( [x_j^{*(v)} + \cos(\delta) x_j^{(*p)}] \right) \right| \le K \sum_j A_{ij} w_{ij} \left| \tanh\left( [x_j^{*(v)} + \cos(\delta) x_j^{(*p)}] \right) \right| \\
     & \le K \sum_j A_{ij} w_{ij} = K
\end{align*}


Given this, when computing the variance and covariance of the opinion distributions, we first normalized the opinions by the parameter $K$. This rescaling ensures that the resulting values fall within the range $[0,1]$. While covariance can, in principle, take negative values, in our case all observed values were positive due to the positive correlations considered in the model.

In contrast, the calculation of the mean conviction was performed on the unnormalized opinion values, as this metric is intended to reflect the agents’ absolute distance from a neutral opinion.

We define an agent’s conviction as the absolute value of their opinion on a given topic, denoted by $|x_i^v|$, where $x_i^v$ is the opinion of agent $i$ on topic $v$. For classification purposes, we compute the mean conviction averaged over all agents and across both topics. Low values of mean conviction indicate that agents' opinions are clustered near zero, suggesting a state of neutral consensus. This metric was specifically used to distinguish between neutral and radicalized consensus states.


The threshold values used for classification were obtained through manual inspection of the data. Next we present the criteria for classification:


\begin{itemize}
    \item If the mean conviction $\langle |x| \rangle < 0.1$, the simulation is classified as a \textit{neutral consensus} state.
    
    \item If the variances in both topics $(\sigma^2_x, \sigma^2_y)$ are below $0.4$, the state is classified as a \textit{radicalized consensus}.
    
    \item If only one of the variances is greater than or equal to $0.4$, the simulation is classified as a \textit{one-dimensional polarization} state.
    
    \item If both variances are greater than or equal to $0.4$, the simulation is considered polarized in both topics. In case the covariance is greater to $0.25$, the state is classified as \textit{ideological alignment}, otherwise, it is classified as \textit{uncorrelated polarization}.
\end{itemize}

\subsection{Mathematical Analysis I: Radicalization}
\label{section: Analisis radicalizacion}

We analyze first the radicalized opinion distribution, in which all agents hold the same opinion $\mathbf{x}^* =(x^{*(1)}, x^{*(2)})$. The equilibrium opinion values are given by the system of equations:
\begin{align}
    x^{*(1)} &= K \tanh\left( x^{*(1)} + x^{*(2)} \cos\delta \right)\; , \\
    x^{*(2)} &= K \tanh\left( x^{*(2)} + x^{*(1)} \cos\delta \right) \; .
\end{align}

\noindent To find the solutions of the system, we impose an ansatz $x^{*(2)}=c x^{*(1)}$, such that both equations result in:
\begin{align}
    x^{*(1)} &= K \tanh\left[ x^{*(1)}(1 + c\cos\delta) \right]\; , \\
    cx^{*(1)} &= K \tanh\left[ x^{*(1)} \left(c + \cos\delta\right) \right] \; .
\end{align}

\noindent There are solutions only for certain values of $c$, which can be found numerically in general. However, in the particular case of $c=1$, both equations reduce to a single one:
\begin{equation}
    x^{*(1)} = K \tanh\left[ x^{*(1)}(1 + \cos\delta) \right]\; ,
\end{equation}

\noindent whose solution is easier to obtain. This nonlinear equation suffers a bifurcation under the condition $1=K(1+\cos\delta)$ such that, if $K<(1+ \cos\delta)^{-1}$, there is a single solution $x^{*(1)}=0$, while if $K>(1+\cos\delta)^{-1}$, there are three possible solutions: $x^{*(1)}=0$, and $x^{*(1)}\equiv \pm x^{*}$, whose values can be found numerically. If $K$ is high enough, then one can obtain that $x^*\simeq K$.

In order to analyze the stability of this fixed point, we linearize the system of equations around $\mathbf{x}^* = (x^*,x^*)$ such that we can express $\dot{\mathbf{x}} = \text{J}\mathbf{x}$, where:
\begin{equation}
    \text{J} = \begin{pmatrix}
\frac{\partial\dot{x}^{(1)}}{\partial x^{(1)}} & \frac{\partial\dot{x}^{(1)}}{\partial x^{(2)}} \\
\frac{\partial\dot{x}^{(2)}}{\partial x^{(1)}} & \frac{\partial\dot{x}^{(2)}}{\partial x^{(2)}} 
    \end{pmatrix}_{(x^*,x^*)} \; ,
\end{equation}

\noindent represents the Jacobian matrix. The system will be stable if conditions $\tau<0$, $\Delta<0$ apply, where $\tau$ denotes the trace of $\text{J}$, and $\Delta$, its determinant. One can easily obtain then the conditions:
\begin{align}
    K&<\cosh^2\left[x^*(1+\cos\delta)\right] \; , \\
    K(1+\cos\delta)&<\cosh^2\left[x^* 
    (1+\cos\delta)\right]\; . \label{eq:stab_sym}
\end{align}

\noindent The second condition is more restrictive than the first one, so it becomes the stability condition for the fixed point. Moreover, as $\cosh(x)> x \; \forall x$, this condition will always be true for $K$ high enough.

The same analysis can be applied if $c=-1$, resulting in a single differential equation:

\begin{equation}
    x^{*(1)} = K \tanh\left[ x^{*(1)}(1 - \cos\delta) \right]\; .
\end{equation}

\noindent We find that the existence condition of such point is $K>(1-\cos\delta)^{-1}$, and the stability conditions follow:
\begin{align}
    K&<\cosh^2\left[x^*(1-\cos\delta)\right] \; , \\
    K(1+\cos\delta)&<\cosh^2\left[x^* 
    (1-\cos\delta)\right]\; .
\end{align}

\noindent Again, the second one is more restrictive than the first, so it becomes the stability condition for the fixed point. When $\cos\delta=0$, it becomes equivalent to the condition for the symmetric points (Eq.~(\ref{eq:stab_sym})); for higher values of $\cos\delta$ the condition becomes more restrictive than the one for the symmetric points,  and for $\cos\delta\to 1$,  both the existence and stability conditions become $K\to \infty$. 

These four points are the only possible stable solutions of the system. To sum up their behavior, for a given value of $\cos\delta$ and $K<(1+\cos\delta)^{-1}$ there is only one stable solution, $x^{*(1)}=0$. For increasing values of $K$, the symmetric points $x^*{(1)} = \pm x^*$ become stable, while the antisymmetric points $x^*{(1)} = \pm x^* = -x^{*(2)}$ remain unstable, and finally for $K$ high enough, all four points become stable.

\subsection{Mathematical Analysis II: Polarization}
\label{section: Analisis polarizacion}

We consider a fully connected population of $N$ agents grouped in $n$ opinion clusters following opinions $\{\mathbf{x_1},...,\mathbf{x_n}\}$. We denote as $N_\xi$ the number of agents holding opinion $\mathbf{x}_\xi$, and we further assume that $N_\xi\gg 1 \; \forall \xi$. Such states are considered to be polarized because of the presence of more than one opinion cluster. The equilibrium opinion for a given agent $i$ belonging to opinion cluster $\xi$ fulfill $\dot{\mathbf{x}}_i = \dot{\mathbf{x}}_\xi = 0$, so that we can express the equilibrium opinions $\mathbf{x}_\xi = (x_\xi^{(1)},x_\xi^{(2)})$ as:
\begin{align}
    x_\xi^{(1)} &= K \sum_{\alpha=1}^{n} N_\alpha \omega_{\xi \alpha} \tanh(x_\alpha^{(1)} + x_\alpha^{(2)} \cos\delta)\; , \label{Eq:xxi1}\\
    x_\xi^{(2)} &= K \sum_{\alpha=1}^{n} N_\alpha \omega_{\xi \alpha} \tanh(x_\alpha^{(2)} + x_\alpha^{(1)} \cos\delta) \; , \label{Eq:xxi2}
\end{align}

Now, to check the stability of such state, we introduce a perturbation $\boldsymbol{\lambda} =\lambda(\cos\varphi,\sin\varphi)$ on an agent $i$ initially holding an opinion $\mathbf{x}_i = \mathbf{x}_\xi$, with $\varphi\in[0,2\pi)$, such that it becomes $\mathbf{x}_i = \mathbf{x}_\xi + \boldsymbol{\lambda}$. We can write her dynamic equations as:
\begin{align*}
    \dot{x}_i^{(1)} = \dot{x}_\xi^{(1)} + \dot{\lambda}\cos\varphi - \lambda\dot{\varphi}\sin\varphi \; , \\
    \dot{x}_i^{(2)} = \dot{x}_\xi^{(2)} + \dot{\lambda}\sin\varphi + \lambda\dot{\varphi}\cos\varphi \; .
\end{align*}

\noindent Considering that $\dot{x}_\xi^{(1)} = \dot{x}_\xi^{(2)}=0$, we can finally obtain that:

\begin{equation}
    \dot{\lambda} = \dot{x}_i^{(1)}\cos\varphi + \dot{x}_i^{(2)}\sin \varphi \; . \label{Eq:derLambda}
\end{equation}

\noindent By computing the value of $\dot{\lambda}$ for all opinion clusters and all possible directions (given by $\varphi$), we can check if the system is stable ($\dot{\lambda}<0$ always) or unstable ($\dot{\lambda}>0$ in some cases). We can further develop the expression considering that:

\begin{equation*}
    \dot{x}_i^{(1)} = -(x_\xi^{(1)}+\lambda \cos \varphi) + K \frac{\sum_{\alpha=1}^{n} N_\alpha \left(d_{i\alpha}+\epsilon\right)^{-\beta} \tanh(x_\alpha^{(1)} + x_\alpha^{(2)} \cos\delta)}{\sum_{\alpha=1}^n N_\alpha \left(d_{i\alpha}+\epsilon\right)^{-\beta}} \; ,
\end{equation*}

\noindent where we have explicitly expressed $\omega_{\xi\alpha}$. Multiplying and dividing by $(d_{i\xi}+\epsilon)^\beta$, we have:

\begin{equation}
    \dot{x}_i^{(1)} = -(x_\xi^{(1)}+\lambda \cos \varphi) + K \frac{\sum_{\alpha=1}^{n} N_\alpha \left(\frac{d_{i\xi}+\epsilon}{d_{i\alpha}+\epsilon}\right)^\beta \tanh(x_\alpha^{(1)} + x_\alpha^{(2)} \cos\delta)}{\sum_{\alpha=1}^n N_\alpha \left(\frac{d_{i\xi}+\epsilon}{d_{i\alpha}+\epsilon}\right)^\beta} \; . \label{Eq:preTaylor}
\end{equation}

\noindent For small values of the perturbation $\lambda$, we can compute the Taylor series expansion:

\begin{equation*}
    \left( \frac{d_{i\xi}+\epsilon}{d_{i\alpha}+\epsilon} \right)^\beta \simeq \left(\frac{\epsilon}{d_{\xi\alpha}+\epsilon}\right)^\beta + \lambda \beta \left(\frac{\epsilon}{d_{\xi\alpha}+\epsilon}\right)^{\beta-1}\left[\frac{\frac{\mathrm{d} d_{i\xi} }{\mathrm{d} \lambda}(d_{i\alpha}+\epsilon) - (d_{i\xi}+\epsilon)\frac{\mathrm{d} d_{i\alpha}}{\mathrm{d}\lambda}}{\left( d_{i\alpha}+\epsilon \right)^2} \right]_{\lambda = 0} \; ,
\end{equation*}

\noindent where we have taken into account that $d_{i\alpha}|_{\lambda=0} = d_{\xi \alpha}$, and $d_{i\xi}|_{\lambda=0} = 0$. We can rewrite this expression as:

\begin{equation*}
    \left( \frac{d_{i\xi}+\epsilon}{d_{i\alpha}+\epsilon} \right)^\beta \simeq f_{\xi\alpha}^\beta + \lambda \beta f_{\xi\alpha}^{\beta-1}g_{\xi\alpha} \; ,
\end{equation*}

\noindent where, explicitly:
\begin{align}
    f_{\xi\alpha} &= \frac{\epsilon}{d_{\xi\alpha}+\epsilon} \; , \\
    g_{\xi\alpha} &= \frac{\sqrt{1+2\sin\varphi\cos\varphi \cos\delta}(d_{\xi\alpha}+\epsilon) +\epsilon \frac{ (x_\alpha^{(1)}-x_\xi^{(1)})(\cos\varphi + \sin\varphi\cos\delta) + (x_\alpha^{(2)}-x_\xi^{(2)})(\sin\varphi + \cos\varphi\cos\delta) }{d_{\xi\alpha}}}{(d_{\xi\alpha}+\epsilon)^2} \; .
\end{align}

\noindent This allows us to rewrite Eq.~(\ref{Eq:preTaylor}) as:
\begin{equation*}
    \dot{x}_i^{(1)} = -(x_\xi^{(1)}+\lambda \cos \varphi) + K \frac{ N_\xi \tanh(x_\xi^{(1)}+x_\xi^{(2)}\cos\delta)+\  \sum_{\alpha\neq \xi} N_\alpha 
    \left( f_{\xi\alpha}^\beta + \lambda \beta f_{\xi\alpha}^{\beta-1}g_{\xi\alpha}\right)\tanh\left(x_\alpha^{(1)} + x_\alpha^{(2)} \cos\delta\right)}{N_\xi + \sum_{\alpha\neq \xi} \left(N_\alpha f_{\xi\alpha}^\beta + \lambda \beta f_{\xi\alpha}^{\beta-1}g_{\xi\alpha}\right)} \; .
\end{equation*}

\noindent The same process can be followed for $\dot{x}_i^{(2)}$, arriving at the expression:
\begin{equation*}
    \dot{x}_i^{(2)} = -(x_\xi^{(2)}+\lambda \sin \varphi) + K \frac{ N_\xi \tanh(x_\xi^{(2)}+x_\xi^{(1)}\cos\delta)+\  \sum_{\alpha\neq \xi} N_\alpha 
    \left( f_{\xi\alpha}^\beta + \lambda \beta f_{\xi\alpha}^{\beta-1}g_{\xi\alpha}\right)\tanh\left(x_\alpha^{(2)} + x_\alpha^{(1)} \cos\delta\right)}{N_\xi + \sum_{\alpha\neq \xi} \left(N_\alpha f_{\xi\alpha}^\beta + \lambda \beta f_{\xi\alpha}^{\beta-1}g_{\xi\alpha}\right)} \; .
\end{equation*}

\noindent The numerator and denominator of the previous equations are linear with $\lambda$, and can be written as:

\begin{equation*}
    \frac{A^{(1,2)}+\lambda \beta B^{(1,2)}}{C + \lambda \beta D} \; ,
\end{equation*}

\noindent where the superindexes $(1,2)$ indicate the corresponding component, and:
\begin{align*}
    A^{(1,2)} &= \sum_{\alpha=1}^n N_\alpha f_{\xi\alpha}^\beta \tanh\left(x_\alpha^{(1,2)} + x_\alpha^{(2,1)}\cos\delta\right) \; , \\
    B^{(1,2)} &=\sum_{\alpha \neq \xi} N_\alpha f_{\xi\alpha}^{\beta-1}g_{\xi\alpha}\tanh\left(x_\alpha^{(1,2)} + x_\alpha^{(2,1)}\cos\delta\right) \; , \\
    C &= \sum_{\alpha=1}^n N_\alpha f_{\xi\alpha}^\beta\; , \\
    D &= \sum_{\xi\neq\alpha}N_\alpha f_{\xi\alpha}^{\beta-1}g_{\xi\alpha} \; .
\end{align*}

\noindent For small values of $\lambda$, we can approximate:

\begin{equation*}
    \frac{A^{(1,2)}+\lambda \beta B^{(1,2)}}{C + \lambda \beta D} \simeq \frac{A^{(1,2)}}{C} + \lambda \beta \frac{B^{(1,2)}C - A^{(1,2)}D}{C^2} \; .
\end{equation*}

\noindent It is easy to see from Eqs.~(\ref{Eq:xxi1}) and (\ref{Eq:xxi2}) that $x_\xi^{(1,2)} = K\frac{A^{(1,2)}}{C}$. Therefore, we can finally express $x_i^{(1)}$, $x_i^{(2)}$ as:
\begin{align}
    \dot{x}_i^{(1)} &= -\lambda \cos \varphi + K\lambda \beta \frac{B^{(1)} C - A^{(1)} D}{C^2}\; . \\
    \dot{x}_i^{(2)} &= -\lambda \sin \varphi + K\lambda \beta \frac{B^{(2)}C - A^{(2)}D}{C^2}\; ,
\end{align}

\noindent and Eq.~(\ref{Eq:derLambda}) as:
\begin{align}
   \dot\lambda &= -\lambda (\cos^2 \varphi + \sin^2\varphi) + K\lambda\beta \frac{(B^{(1)}C-A^{(1)}D)\cos\varphi + (B^{(2)}C-A^{(2)}D)\sin\varphi}{C^2} = \nonumber \\
   &= \lambda \left[ K\beta \frac{(B^{(1)}C-A^{(1)}D)\cos\varphi + (B^{(2)}C-A^{(2)}D)\sin\varphi}{C^2} - 1 \right] \; . \label{Eq:derLambda2}
\end{align}

\begin{figure}[t]
    \centering
     \includegraphics[width=\linewidth]{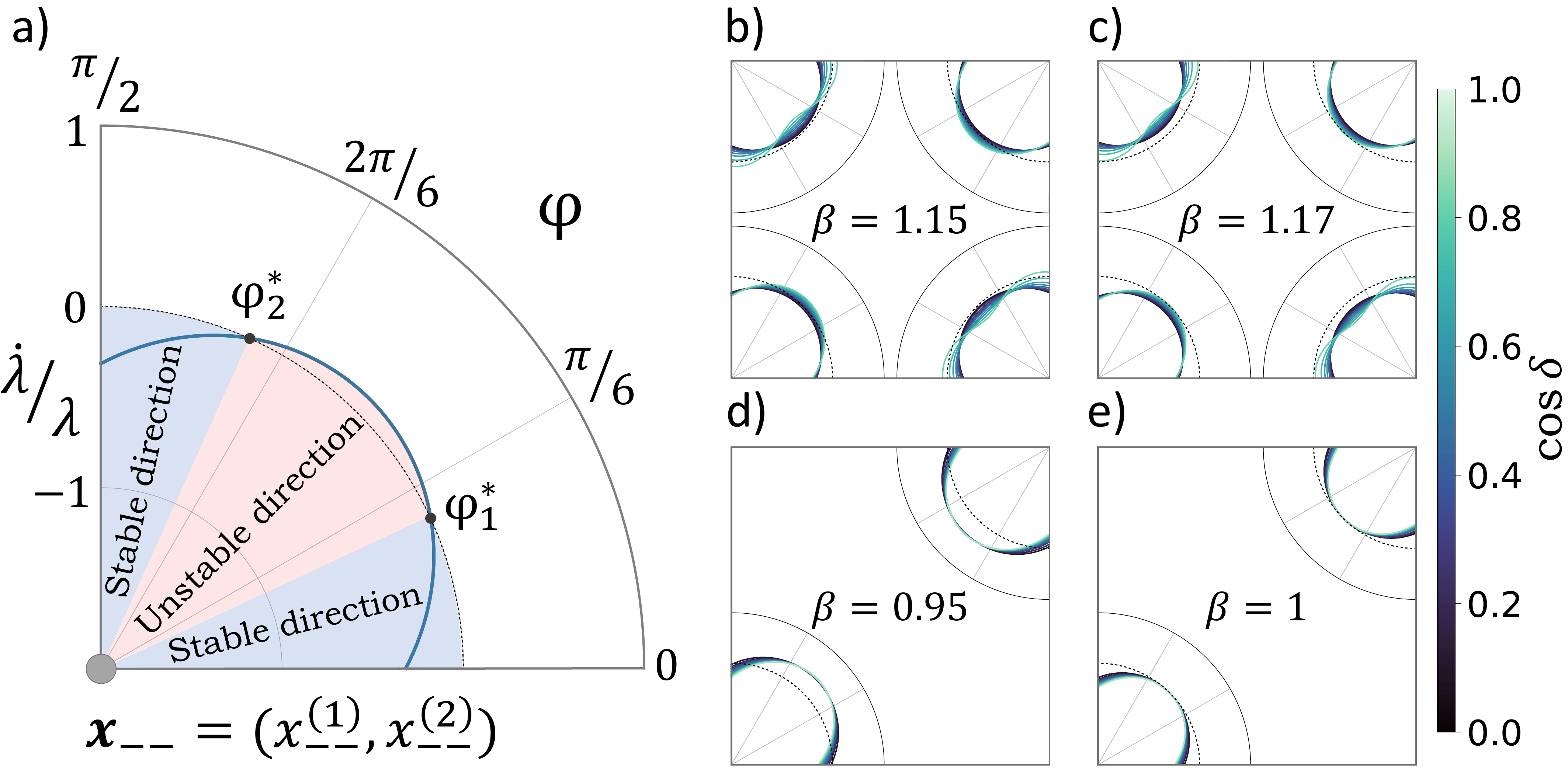}
    \caption{Diagrams for the stability condition given by Eq.~(\ref{Eq:stab_cond}). (a) Example for $\mathbf{x}_{--}$ at $\beta=1.14$ and $\cos\delta=0.3$. The angle represent the perturbation direction $\varphi$, and the radius, the relative change in value of the perturbation modulus, $\dot\lambda/\lambda$. For values of $\varphi$ close to $0$ or $\pi/2$ the perturbation is stable, but it remains unstable for $\varphi$ close to $\pi/4$. The transition occurs at angles $\varphi_1^*$ and $\varphi_2^*$. Therefore, the opinion cluster $\mathbf{x}_{--}$ is unstable under perturbation. (b) and (c) Stability of the four opinion clusters corresponding to uncorrelated polarization for (b) $\beta=1.15$ and (c) $\beta=1.17$. (d) and (e) Stability of the two opinion clusters corresponding to ideological polarization for (d) $\beta=0.95$, and (e) $\beta=1$. Throughout the figure, $K=10$ and $N_\alpha\simeq N/n, \, \alpha=1,...,n$.}
    \label{fig:stab_dir}
\end{figure}

\noindent Therefore, the system will be stable if the following condition

\begin{equation}
    K\beta \frac{(B^{(1)}C-A^{(1)}D)\cos\varphi + (B^{(2)}C-A^{(2)}D)\sin\varphi}{C^2} < 1 \; , \label{Eq:stab_cond}
\end{equation}

\noindent holds, which depends on parameters $\beta$ and $K$, and the direction of the perturbation $\varphi$. In Fig.~\ref{fig:stab_dir}a we show a diagrammatic example that summarizes the expected behavior of an agent holding a given opinion $\mathbf{x}_{--}$: depending on $\varphi$, the system can be either stable ($\dot\lambda /\lambda<0$) or unstable ($\dot\lambda /\lambda>0$) under the proposed perturbation for given values of $\beta$ and $\cos\delta$. The opinion cluster will only be stable under perturbation if it is so for all possible directions.

We now consider the two most relevant polarized states that appear in the system, uncorrelated polarization and ideological polarization. In the following, we assume that $N_\alpha\simeq N/n,\; \alpha=1,...,n$, and compute the value of $\dot\lambda/\lambda$ following Eq.~(\ref{Eq:derLambda2}), representing its result following the example of Fig.~\ref{fig:stab_dir}a. In the case of uncorrelated polarization, we must consider four opinion clusters $\{\mathbf{x}_{++},\mathbf{x}_{+-},\mathbf{x}_{-+}, \mathbf{x_{--}}\}$, whose values are given by solving Eqs.~(\ref{Eq:xxi1}) and (\ref{Eq:xxi2}). These clusters are located near the four edges of the opinion plane $\{x^{(1)},x^{(2)}\}$. In Figs.~\ref{fig:stab_dir}b and \ref{fig:stab_dir}c we represent $\dot\lambda/\lambda$ for all opinion clusters and for different values of $\beta$ and $\cos\delta$, as a function of the perturbation direction. Evidently, the behavior for $\mathbf{x}_{++}$ and $\mathbf{x}_{--}$ is the same due to the system's symmetry, as is the behavior for $\mathbf{x}_{+-}$ and $\mathbf{x}_{-+}$. For $\beta=1.15$ (Fig.~\ref{fig:stab_dir}b), if $\cos\delta$ is low enough, the system is stable under any perturbation. However, for higher values of $\cos\delta$, there are multiple directions in which $\dot\lambda/\lambda>0$ rendering the system unstable. On the contrary, for $\beta=1.17$ (Fig.~\ref{fig:stab_dir}c), the system becomes stable under all directions of the perturbation and all the equilibrium points for all values of $\cos\delta$. 

The same behavior can be inferred for ideological polarization, in Figs.~\ref{fig:stab_dir}d and \ref{fig:stab_dir}e, where we consider two possible opinion clusters holding $\{\mathbf{x}_{++},\mathbf{x}_{--}\}$. Again, the behavior is similar for both clusters thanks to the system's symmetry, and it can be easily verified that $x_i^{(1)}=x_i^{(2)}\equiv x_i$ $\forall i$, placing all agents over the system's diagonal. If $\beta<1$ (Fig.~\ref{fig:stab_dir}d), the opinion clusters are unstable under perturbations in the general direction of the opposed opinion cluster. However, for $\beta=1$ (Fig.~\ref{fig:stab_dir}e), the system becomes stable under all possible perturbations for all values of $\cos\delta$ at the same time. This result is a direct consequence of the system becoming effectively one-dimensional in the case of ideological polarization, as we argue on the main text. Under these circumstances, $A^{(1)}=A^{(2)}$ and $B^{(1)}=B^{(2)}$, so the equilibrium condition from Eq.~(\ref{Eq:stab_cond}) becomes:

\begin{equation}
        (\cos\varphi + \sin\varphi)K\beta \frac{B^{(1)}C-A^{(1)}D}{C^2} < 1 \; , \label{Eq:stab_cond_2}
\end{equation}

\noindent In the particular case of $\varphi=\pi/4$, the stability condition can be expressed as:

\begin{equation}
     A\left( \frac{\epsilon}{\epsilon + |x_{++}-x_{--}|\sqrt{2(1+\cos\delta)}} \right)^{\beta-1} < 1 \; , \label{Eq:equilibrium_ideol}
\end{equation}

\noindent where, explicitly:
\begin{equation}
    A = \sqrt{2} K\beta N_{++}N_{--} \frac{\tanh[x_{++}(1+\cos\delta)]-\tanh [x_{--}(1+\cos\delta)]}{\left[N_{--} + N_{++}\left(\frac{\epsilon}{\epsilon + |x_{++}-x_{--}|\sqrt{2(1+\cos\delta)}}\right)\right]^2}\frac{(1+\cos\delta)\left(\sqrt{2}|x_{++}-x_{--}| + \epsilon\right)}{\left[\epsilon + |x_{++}-x_{--}| \sqrt{2(1+\cos\delta)}\right]^2} \; .
\end{equation}

\noindent The inclusion of the second opinion axis results in new factors $1+\cos\delta$ and $\sqrt{2}$ appearing on the stability condition with respect to the one-dimensional case (see~\cite{Pol_opinions_2023}). Regardless, it can be noticed that the numerator of $A$ is proportional to $8K^2 \beta N_{++}N_{--}(1+\cos\delta)$, considering that $\epsilon\ll|x_{++}-x_{--}|$ and, if $K$ is high enough, $|x_{++}-x_{--}|\simeq 2K$ and $\tanh[x_{++}(1+\cos\delta)] \simeq \tanh(x_{++}) \simeq 1$. Moreover, the denominator of $A$ is proportional to $N_{--}^2[2K\sqrt{2(1+\cos\delta)}]^2 = 8K^2N_{--}^2(1+\cos\delta)$. Therefore, we conclude that $A\simeq\beta$ regardless of the value of $\cos\delta$. Finally, the term that accompanies $A$ in Eq.~(\ref{Eq:equilibrium_ideol}) becomes $1$ for $\beta=1$, and thus, we find that $\beta=1$ becomes the transition condition for all values of $\cos\delta$, and the perturbation direction $\varphi=\pi/4$. A similar argument can be applied for $\varphi=5\pi/4$, arriving at the same result.

\begin{figure}[t]
    \centering
     \includegraphics[width=.9\linewidth]{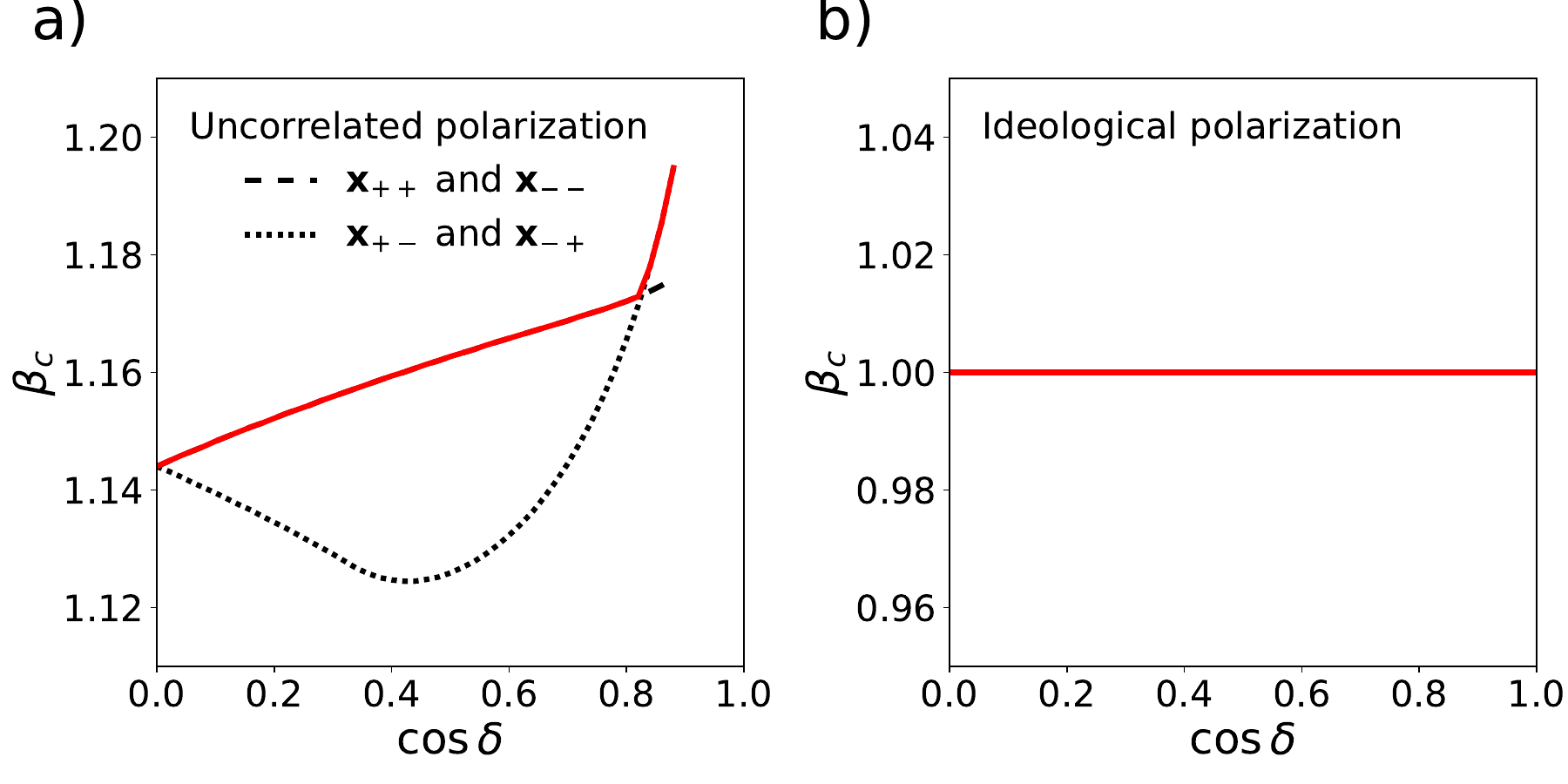}
    \caption{Values of $\beta_c$ as a function of $\cos\delta$. (a) $\beta_c$ for uncorrelated polarization: for $x_{++}$ and $x_{--}$ (dashed black line), for $\mathbf{x}_{+-}$ and $\mathbf{x}_{-+}$ (pointed black line), and for the whole state (solid red line). (b) Ideological polarization: $\beta_c$ for ideological polarization (solid red line). Throughout the figure, $K=10$ and $N_\alpha\simeq N/n, \, \alpha=1,...,n$.}
    \label{fig:stab}
\end{figure}

We summarize the obtained values of $\beta_c$ in Fig.~\ref{fig:stab}, where we show the computed critical value for each equilibrium point as a function of $\cos\delta$. The condition for global stability will be that value of $\beta$ such that all points are stable. In the case of uncorrelated polarization (Fig.~\ref{fig:stab}a), the behavior for points on the favored ideological axis and outside of it is different: in general, $\mathbf{x}_{+-}$ and $\mathbf{x}_{-+}$ become stable for lower values of $\beta$ than $\mathbf{x}_{++}$ and $\mathbf{x}_{--}$, but this behavior changes for very high values of $\cos\delta$ due to the increased instability of points outside the diagonal that eventually leads to its disappearance before reaching $\cos\delta=1$. In the case of ideological polarization (Fig.~\ref{fig:stab}b), as we have already explained, the transition occurs at $\beta=1$ consistently. Therefore, we conclude that ideological polarization is invariably more stable than uncorrelated polarization, and thus, the presence of a larger number of ideologies makes polarization more difficult.

\subsection{The ANES survey}
\label{section: Encuesta}

The American National Election Studies (ANES) is a long-standing survey performed nation-wide across the US populace that analyzes their voting behavior and political opinions during presidential elections regarding multiple topics before and after the electoral process. This is a free access data, greatly used not only in scientific publications, but also as a source for books or news articles, counting with over 9600 citations.

We extract the opinions' distributions about eleven topics from the ANES 2020 dataset, the most recent presidential election during the time in which we started our work. The topics considered are the following:

\begin{itemize}
    \item \textit{President didn't have to worry about Congress:} Would it be helpful, harmful, or neither helpful nor harmful if U.S. presidents could work on the country’s problems without paying attention to what Congress and the courts say? (code: V201372x)
    \item \textit{Impeachment:} Do you favor, oppose, or neither favor nor oppose the U.S. House of Representatives’ decision in December of last year to impeach President Trump? (code: V201386x)
    \item \textit{Service to same sex couples:} Do you think business owners who provide wedding-related services should be allowed to refuse services to same-sex couples if same-sex marriage violates their religious beliefs, or do you think business owners should be required to provide services regardless of a couple’s sexual orientation? (code: V201408x)
    \item \textit{Transgender policy:} Should transgender people - that is, people who identify themselves as the sex or gender different from the one they were born as - have to use the bathrooms of the gender they were born as, or should they be allowed to use the bathrooms of their identified gender? (code: V201411x)
    \item \textit{End birthright citizenship:} Some people have proposed that the U.S. Constitution should be changed so that the children of unauthorized immigrants do not automatically get citizenship if they are born in this country. Do you favor, oppose, or neither favor nor oppose this proposal? (code: 201420x)
    \item \textit{Wall with Mexico:} Do you favor, oppose, or neither favor nor oppose building a wall on the U.S. border with Mexico? (code: V201426x)
    \item \textit{Less or more government:} Would it be good for society to have more government regulation, about the same amount of regulation as there is now, or less government regulation? (code: V202255x)
    \item \textit{2010 health care law:} Do you approve, disapprove, or neither approve nor disapprove of the Affordable Care Act of 2010, sometimes called Obamacare? (code: V202328x)
    \item \textit{Requiring vaccines in schools:} Do you favor, oppose, or neither favor nor oppose requiring children to be vaccinated in order to attend public schools? (code: V202331x)
    \item \textit{Background checks for gun purchases:} Do you favor, oppose, or neither favor nor oppose requiring background checks for gun purchases at gun shows or other private sales? (code: V202341x)
    \item \textit{Government action about opioids:} Do you think the federal government should be doing more about the opioid drug addiction issue, should be doing less, or is it currently doing the right amount? (code: V202350x)
\end{itemize}

The chosen questions correspond to different periods of surveying, before and after election. In particular, issues "Less or more government'', "2010 health care law'', "Requiring vaccines in schools'', "Background checks for gun purchases'' and "Government action about opioids'' correspond to the post-election survey, and the rest, to the pre-election survey. The set of eleven questions was selected with the intent of having a mix of bimodal and unimodal distributions of opinions, which would represent polarized and consensus states respectively. Populations for both surveys (pre and post election) vary slightly, and thus, different weights must be used to extrapolate the results to the US population. Weights for the pre-election survey correspond to code V200010a, and for the post-election survey, to code V200010b. Our data was constructed combining the opinion distributions of pair of topics. In the case both questions belonged to the pre-election survey, the weights to be used would be the ones corresponding to code V200010a. In any other case, the weights to be used would be the ones corresponding to V200010b. This is as instructed in the ANES guidebook.

Opinion histograms are normalized removing all missing data, comprised of answers like "don't know'', "refused to answer'', survey errors or inexistence of survey data, and also neutral answers like "neither favor nor oppose" or other similar answers equidistant from the two extreme opinions.

\end{document}